\documentstyle[twocolumn,aps,floats]{revtex}

\begin{document}

\draft \tolerance = 10000

\setcounter{topnumber}{1}
\renewcommand{\topfraction}{0.9}
\renewcommand{\textfraction}{0.1}
\renewcommand{\floatpagefraction}{0.9}
\newcommand{\br}{{\bf r}}

\twocolumn[\hsize\textwidth\columnwidth\hsize\csname
@twocolumnfalse\endcsname

\title{Is the Time a Dimension of an Alien Universe?}
\author{L.Ya.Kobelev \\
Department of  Physics, Urals State University \\ Lenina Ave., 51,
Ekaterinburg 620083, Russia  \\ E-mail: leonid.kobelev@usu.ru}

\maketitle

\begin{abstract}
On the base of the hypothesis about a nature of the time as a dimension of
alien Universe  relation between alteration of  time with coordinates
$\frac{\partial t}{\partial x}$ and time {t} offered: $ \frac{\partial t}
{\partial x} = H_{t} t$ . This relation is an analogy of the Habble law in
the time space. The consequence of it is  additional redshift $Z_{DT}$
depending on differences $\tau$ of times  existence  of the objects with
redshift that  are compared ($t_{0}$ is the time existence of more old
object):
$Z_{DT}=\frac{1+\frac{\tau}{t_{0}}}{\sqrt{1-(\frac{\tau}{t_{0}}})^{2}}-1$.
The redshift of Arp galaxies may be explained if this relation is used and
this explanation doe's not contradict Arp hypothesis about supernova
explosions. Discussion a possibilities of experimental verification  of
the hypothesis is considered.
\end{abstract}

 \vspace{1cm}

]

\section {Introduction}

    The appearance in recent years many variants of quantum gravity, including
theories with change of signature sign, gravitational quantum tunnel
transitions in the inflation phase, different methods of quantizing and so
forth (see, e.g., a review in \cite{alt}) do not allow  to choose the
criteria to distinguish "right" theories from "wrong" ones. The situation
is similar to that existed in the beginning of our century in physics,
when Lee groups theory was invented and included as a special case the
Lorenz sub-group but there were no convincing reasons to single out of it
the Dirac equation. Now there is no a quite clear idea on the time nature
which would permit to predict the phenomena whose experimental examination
in its turn would help to judge its reliability. If there is any, it would
be able to significantly narrow down the choice for possible quantum
gravity theories. It is one of such "time models" that is discussed in the
present paper. Treating time as a dimension of another universe with laws
similar to those of our world, it gives the way to a number of
experimentally phenomena, some of which are discussed in the paper.

\section {Intersecting universes hypothesis}

    Let us consider time to be one of dimensions of a three-dimensional space
of another universe which adjoins to ours and penetrate into it by this
dimension (see \cite{kob}). To do this, assume that at the "Big Bang" the
quantum state of the early universe gave the birth to an ensemble of
inflation formations with different values of Hubble constant
$H=\frac{\dot{a}}{a}$ ($a$ is the scale factor of the space elements of
metric). This does not contradict to the contemporary conception of the
world in the inflation universe models. Then a chains of initially
three-dimensional formations also could emerge, with one dimension
penetrating into another one and thus making it four-dimensional. So,
restrict ourselves to only such formations. Stress, that this formations
are primordial three-dimensional but in addition there exist a
"penetration" of one dimension of another formation (we shall call it
"neighboring") into it. In other words, we are interested in only the
formations having total number of dimensions being equal to four:  three
of them are "own" and the fourth one penetrated from another formation and
is of different, generally speaking, origin. Thus, this fourth dimension
belongs to both of the neighboring formations and is common to them. It is
"own" to one formation and "foreign" to another. We shall refer to such
formations as "universes".
    In the neighboring universe (whose the fourth dimension is)
the situation would be the same. To its own three dimensions (one of which
is common with the first universe considered) a new "foreign" dimension
from another universe is added - either from that mentioned before or not.
In the first case we  have an isolated pair of universes, and in the
second one have the chain of intersecting universes which can be closed
and finite or open and infinite.
    If we are to suppose non-equal inflating of the three-dimensional volumes of
this universes in regard to each other (because of different value $a$ ),
the fourth dimension penetrated into a universe from another one can play
the role of natural calibrating value while comparing the volumes and be
treated as "time". In this case the dimension of "foreign" universe common
to ours is time to us, but in the foreign universe the role of time is
played by a dimension of an alien to it universe (either new one, or,
probably, ours). Note, that this model does not reduce to the York  theory
(see \cite{york}) or to the "scale" theories of time. Here time is not
simply a scale factor but a dimension of origin and properties different
from that of our three space dimensions, though it plays in its own
universe role exactly the same as them.
    All what is needed for such treatment is a difference in expanding
between the three "own" and the fourth "foreign" dimensions. Therefore,
from this point of view, time can be considered as a dimension of another
formation and common to our universe but with properties different from
that of our own three spatial dimensions and, hence, expanding unlike
them. The concept of the expansion rate of the universe (as well as time
and rate concepts themselves) are then defined by comparing the laws of
changing three-dimensional volumes of neighboring universes having one
common (the fourth to us) dimension. The four-dimensional interval of
special relativity $ds^{2}=dr^{2}-c^{2}dt^{2}$ in this case, at $ds^{2}=0$
will yield the rate of changing of our spatial coordinates in relation to
$t$ coordinate which characterize the alteration of the "spatial"
dimensions of the other universe. Light velocity $c$ in this terms is
simply the difference between changing in own three-dimensional volumes of
two neighboring universes. It appeared as a result of the Big Bang and can
be exceeded in our space only at local but not at large scales. In the
neighboring universe, provided validity after the phase of blowing a kind
of special relativity condition like that of our world, we shall have
$ds^{*2}=dt^{2}-c^{*2}dr^{2}$, and, $c^{*}=c^{-1}$ from our point of view,
in the simple case. What consequences arise from this hypothesis
concerning the nature of time, based on existing of at least two such
four-dimensional universes with one common dimension?

\section {Anomalous Redshift Behavior}

    As the first consequence of the model proposed we shall discuss the red
shift in the spectra of far star objects. Let the physical laws of nature
be equal in the neighboring universes. Then in both of them, ours
"spatial" and the other "temporal" the following Hubble laws can be
written:
\begin{equation}\label{eq1}
 \frac{dx}{dt}=v_{x}=H_{x}x
\end{equation}
\begin{equation}\label{eq2}
\frac{dt}{dx}=v_{t}=H_{t}t
\end{equation}

where $v_{x}$ and $v_{t}$ are the rates of scattering of star objects in
the corresponding universes, $H_{x}$ and $H_{t}$ are the Hubble constants
in them and $x$ and $t$ mean the distances between the objects in the
"spatial" and "temporal" universes. One of three dimensions making in the
"temporal" universe vector $t$ penetrates into our universe and plays the
role of time to us.We shall denote it as $t$, because only one component
of $t$ is common. Stress, that instead of the Hubble law (\ref{eq1}) now
one should take into account in addition to it the new relation
(\ref{eq2}), that is

\begin{equation}\label{eq3}
v_{t}=H_{t}t
\end{equation}

The physical sense of $v_{t}$ is the rate of time alteration per distance
unit and can be bound up with the time passed since the star object was
born in our universe. The Doppler wavelength shift
$Z=(\lambda_{0}-\lambda_{e})/\lambda_{e}$ (here $\lambda_{0}$ and
$\lambda_{e}$ are wavelengths in the observer's frame of reference and in
the object's own one) then should be calculated using (\ref{eq3}) as well
as (\ref{eq1}). Denote it as $Z_{DT}$ and $Z_{DX}$ accordingly. If
$Z_{DT}\ll Z_{DX}$ or $Z_{DT}\gg Z_{DX}$ then the resulting shift is
simply their sum:
\begin{equation}\label{eq4}
Z=Z_{DX}+Z_{DT}
\end{equation}

For $Z_{DT}$, using (\ref{eq3}) one easily finds
\begin{equation}\label{eq5}
Z_{DT}=\frac{1+\frac{v_{t}}{c^{*}}}{\sqrt{1-\frac{v_{t}^{2}}{c^{*2}}}}-1
\end{equation}

where $c^{*}=c^{-1}$. When $c^{*}=\alpha c^{-1}$, $\alpha =const$. Putting
$v_{t}$ from (\ref{eq3}) into (\ref{eq5}) and assuming, in virtue of
similarity of the nature's laws in both of the universes,
$H_{t}=H_{0}c^{-1}$ we obtain (for $c^{*}=c^{-1}$)
\begin{equation}\label{eq6}
v_{e}c^{*-1}=H_{t}t c^{*-1}=H_{0}t(c^{*}c)^{-1}=\frac{t}{t_{0}}
\end{equation}

with $t_{0}=H_{0}^{-1}$ being the time since the universe formation and
$t$ is the difference between the birth time of the star objects of whose
spectra redshift is investigated (i.e, $t=t_{e}-t_{0}$ or $t=t_{0}-t_{e}$
depending on the sign of the expression). Then
\begin{equation}\label{eq7}
Z_{DT}=\frac{1+\frac{t}{t_{0}}}{\sqrt{1-(\frac{t}{t_{0}}})^{2}}-1
\end{equation}

and
\begin{equation}\label{eq8}
Z=Z_{DT}+Z_{DX}\approx\frac{1+\frac{v}{c}}{\sqrt{1-(\frac{v}{c})^{2}}}+
\frac{1+\frac{t}{t_{0}}}{\sqrt{1-(\frac{t}{t_{0}})^{2}}}-2
\end{equation}

If the contributions of $Z_{DT}$  and $Z_{DX}$  into $Z$ are comparable by
value, it is not clear which way to calculate the Doppler shift in the
direction to the observer is more preferable,
\begin{eqnarray}\label{equ1} \nonumber
  Z=Z_{DX}+Z_{DT}
\end{eqnarray}
or
\begin{equation}\label{eq9}
       Z=\frac{1+A}{\sqrt{1-A^{2}}}-1
\end{equation}

where
\begin{eqnarray}\label{equ2} \nonumber
A=\frac{\frac{v_{x}}{c}+\frac{t}{t_{0}}}{1+\frac{v_{x}t}{ct_{0}}}
\end{eqnarray}

The first case corresponds to independent contribution of $Z_{DT}$ and
$Z_{DX}$, while (\ref{eq9}) represents the attempt to add velocities of
different origin (in different universes) in the direction to the observer
(note, that velocity in the "temporal" universe is a vector only in that
universe, but not in ours).

\section{New Energy -Momentum relation for Alien Non-Relativistic Particles?}

As the second example we put the following question: whether a particle
(with rest mass $m_{0}$) can penetrate from one universe into another
(say, from ours to the neighboring or into our universe from elsewhere)
and if yes, which energy it should posses? To answer it, let first write
two Dirac equations for a free particle in our universe and in the other
one:
\begin{equation}\label{eq11}
\partial_{\gamma}\psi(x)=0;               (x=r, it)
\end{equation}
\begin{equation}\label{eq12}
\partial_{\gamma}\psi(x_{t})=0;                (x_{t}=t, ix)
\end{equation}

where $\gamma $ means Dirac matrices. Energy eigenvalues for the particle
then

\begin{equation}\label{eq13}
E^{2}=p^{2}+E_{0}^{2};  E_{0}^{2}=m_{0}c^{2}
\end{equation}
\begin{equation}\label{eq14}
P_{t}^{2}=E_{t}^{2}c^{*2}+P_{0t}^{2}; P_{0}=m_{0}c^{*}
\end{equation}

    Here the particle energy and momentum are written for our and the
neighboring universes accordingly. Note, that momentum corresponds to the
space derivative of wave function, that is $\partial \Psi/\partial r$ and
energy corresponds to the time derivative $\partial \Psi/\partial t$. As
space and time exchange their role while turning from one universe to the
other, the derivatives also will exchange their role. That is, what we
usually call "momentum" will behave like energy and vice versa. We shall
keep the notations $P$ for $\partial \Psi/\partial r$ and $E$ for
$\partial \Psi/\partial t$ though their sense in different universes may
change and this fact is taken into account in (\ref{eq14}).

    If we may admit that when a particle tunneling from one universe into
another its momentum remains, then $P_{t}=P$ and (\ref{eq13}) takes the
form
\begin{equation}\label{eq15}
E^{2}=(E_{t}^{2}c^{*2}+P_{0t}^{2})c^{2}+E_{0}^{2}
\end{equation}

But in the neighboring universe $P_{0t}=m_{0}c^{*}$ and in our universe
momentum can be of any quantity. Therefore $E^{t}$ turns into relativistic
energy and can not be less then $E_{0}=m_{0}c^{2}$. Then from (\ref{eq15})
follows an inequality for the particle energy
\begin{equation}\label{eq16}
E> 2E_{0}
\end{equation}

    This condition is necessary but not sufficient, because our assumptions
require that vector momentum conservation law must realize. Probably, the
validity of this law can be achieved only for a selected direction in our
space or only near massive bodies which get the remainder energy and
momentum. Nevertheless, the expression (\ref{eq16}), in principle, can be
examined experimentally, since it indicates on the possibility of an
apparent violations of the laws of conservation of energy, charge, number
of particles faster than $\sqrt{3/4c}$, which originates from the
particles going out into another universe. For massless particles (i.g.,
photons) this transition may reduce to a frequency renormalization (a
simultaneous for all the frequencies "reddening"). Beside that, one can
take into consideration a possibility for both mass and massless particles
 tunneling through a potential barrier of unknown nature which divided he
universes since they was born.
      Discuss now some aspects of tunneling of quantum particles from one
universe into another. If it is possible, no matter how small the
probability of such a transition is, it seems that one can point out two
observable consequences. The first of them is as follows. In the
neighboring universe a free non-relativistic particle obeys the equation
\begin{equation}\label{eq17}
ih\frac{\partial}{\partial
x}\Psi=\frac{h^{2}}{2m_{0}c^{3}}\frac{\partial^{2}}{\partial t^{2}} \Psi
\end{equation}

therefore the relation between its energy and momentum will differs from
that for a particle in our universe. Namely,
\begin{equation}\label{eq18}
E=\sqrt{2E_{0}c} \sqrt {p}
\end{equation}

\section{Does Intensity of Relic Radiation have a Second Maximum? }

The second consequence observable appears from the relic radiation of the
neighboring universe. This radiation, if tunneling through the barrier
between the universes, can, in principle, be detected. Its frequency and
wavelength can be calculated using the assumed earlier similarity of the
physical laws in both of the universes. Thus, in the neighboring one the
relic radiation wavelengths $\lambda _{tr}$ (measured in seconds) will
coincide numerically with the wavelength obtained in our universe $\lambda
_{r}$. But while turning to our universe from the "foreign" one, the
wavelength must be multiplied by the light velocity $c$. So, $\lambda
_{r}$ of the foreign relic radiation came to us is ( in case
$c^{*}=c^{-1}$)
\begin{equation}\label{eq19}
\lambda _{rt}=\lambda {c}
\end{equation}

Though in (\ref{eq19}) the wavelength is very large (it corresponds to the
frequencies of about 0,1 $sec^{-1}$), at the intensity curve of relic
radiation background there will be an anomaly in this (or another) region,
observed as a maximum with fast decreasing low frequency border. Whether
it is possible to distinguish this maximum from the galaxies heat noise?

\section{Is the Redshift of Arp Galaxies  Consequence of
Differences in Time Existence of Their Parts? }

Very interesting problem is interpretation of large differences in value
of redshift of different parts of Arp galaxies\cite{arp}. All difficulties
may be take of if use for explaining this the new relation (equation (7))
The hypothesis of Arp about explosion of supernova stars may be includes
in this interpretation . The interpretation of redshift is: the different
parts of Arp galaxies were borne in different times.

\section {Final remarks}

    In the frames of the model considered a number of other questions arises.
Is our universe non-isotropic in relation to the direction of "foreign"
dimension penetrating? Is there only a pair of universes or every
dimension can penetrate into some universe or other making an infinite
series of universes? If it is possible to send a particle into another
universe in what time it will find itself when returned and so on.
Apparently answering such questions can stimulate producing more
complicated theories (including the proposed here as a special case) and
further investigations the nature of time.\\\\
    In conclusion we name again the main ideas and results.

    1. The model of the "Universes" discussed treats the time as one of the
dimensions of a "neighboring" Universes born together with ours and
expanding by another law. The rate of expanding is calibrated by comparing
the change in "own" three-dimensional volumes of the neighboring
universes.

    2. If we are to assume the similarity of the physical laws in
different universes, then beside the Hubble law the analogous law is exist
concerning to the rate of changing the time coursing in dependence on the
time passed. This law allows to avoid some contradictions related to
redshift in spectra of a number of anomalous far galaxies (including Arp
galaxies).

    3. The model considered gives way to a number of statements which can
in principle be examined by  experiment. Some of the consequences are
briefly discussed: existing of relic radiation come from "foreign"
universes; a possibility for non-conservation of energy, charge and number
of particles with energies greater than $2E_{0}$ (because of leaving into
the neighboring universes); anomalous dependence of energy upon momentum
for nun-relativistic particles of "foreign" origin.

    All the results following from the model of Universe structure
proposed, though it seems to be unlikely and leading to an unusual
treatment of time, allow to hope for an experimental examination.
Stress,the hypothesis about nature of time doe's non contradict the
hypothesis of multifractal nature of time and space  presented by author
in \cite{kob1}-\cite{kob9}. Some results of multifractal theory of time
and space coincide with results of hypothesis considered in this paper,
but here was presented hypothesis that  gives new look on the possible
origin of time and its nature. As the last remark we once more pay
attention on possibility to receive the results concerning the redshift if
simple postulate the inhomogeneity of time flows and relation $
\frac{\partial t} {\partial x} = H_{t} t$ without any explanations.

\end{document}